\documentclass[aps,preprint,prd,showpacs,nofootinbib]{revtex4}
\usepackage{latexsym}
\usepackage{amsmath}
\usepackage{graphicx}
\usepackage{subfigure}
\usepackage{dcolumn}
\usepackage{bm}
\usepackage{amssymb}
\usepackage{latexsym}
\usepackage{ulem}
\usepackage{color}
\usepackage[colorlinks,linkcolor=magenta,anchorcolor=cyan,citecolor=blue]{hyperref}

\def\be{\begin{equation}}
\def\ee{\end{equation}}
\def\ba{\begin{eqnarray}}
\def\ea{\end{eqnarray}}

\begin{document}

\title{Massive vector particles tunneling from Kerr and Kerr-Newman black holes}

\author{Xiang-Qian Li$^{1}$\footnote{lixiangqian13b@mails.ucas.ac.cn (Corresponding author)}}
\author{Ge-Rui Chen$^{2}$}

\affiliation{$^1$ School of Physics, University of Chinese Academy of
Sciences, Beijing 100049, China}

\affiliation{$^2$ College of Science, Central South University of Forestry and Technology, Changsha 410004, China}

\begin{abstract}

In this paper, we investigate the Hawking radiation of massive spin-1 particles from 4-dimensional Kerr and Kerr-Newman black holes. By applying the Hamilton-Jacobi ansatz and the WKB approximation to the field equations of the massive bosons in Kerr and Kerr-Newman space-time, the quantum tunneling method is successfully implemented. As a result, we obtain the tunneling rate of the emitted vector particles and recover the standard Hawking temperature of both the two black holes.
\end{abstract}
\maketitle

\section{Introduction}

Hawking proved that black hole can release radiation thermodynamically, due to quantum vacuum fluctuation effects near the event horizon \cite{hawking1,hawking2}. Since then, several early methods were proposed to derive the Hawking radiation \cite{tdr,gwgs,bndd,fulling,frolov,wald}, mostly relying on quantum field theory on a fixed background. Later, a semiclassical derivation of Hawking radiation as a tunneling process, including null geodesic method and Hamilton-Jacobi method, has been developed and has attracted a lot of attention. The tunneling methods provided people an alternative way
of understanding black hole radiation. Both approaches to tunneling used the fact that the tunneling probability for the classically forbidden trajectory from inside to outside the horizon is given by $\Gamma={\rm exp}\left(-2{\rm Im}I/\hbar\right)$, where $I$ is the classical action of the trajectory. The difference between these two methods consists in how the imaginary part of the classical action is calculated.

Null geodesic method was first put forward by Kraus and Wilczek \cite{pkfw1,pkfw2} and then developed by Parikh and Wilczek \cite{mkpf,mkp1,mkp2}. For this method, the only part of the action that contributes an imaginary term is $\int_{r_{in}}^{r_{out}}p_{r}dr$, where $p_{r}$ is the momentum of the emitted null s-wave. One can compute the imaginary part of the action by using Hamilton¡¯s equation and knowledge of the null geodesics. Works based on null geodesic method consist of \cite{ecv,ajmm,maam,zzhao1,zzhao2}.

Hamilton-Jacobi method was proposed by Angheben \emph{et al}. \cite{mamn}, as an extension of Padmanabhan's works \cite{kstp,sskt}. This method applies the WKB approximation to the Klein-Gordon equation, with the Hamilton-Jacobi equation as the lowest order. Then according to the symmetry of the metric, one can pick an appropriate ansatz for the action and put it into the Hamilton-Jacobi equation to solve. Applying this method, people studied the Hawking radiation of various space-times and made a great deal of successes~\cite{rkrbm,ajmme,pmit,dysz}. In a way similar to Hamilton-Jacobi method, R. Kerner and R.B. Mann \cite{kern1,kern2} applied the WKB approximation to the Dirac equation to calculate Dirac particles' Hawking radiation.

Kerr and Kerr-Newman metrics describe the geometry of vacuum space-time around a rotating uncharged and charged axially-symmetric black holes, both of which are of great astrophysical significance or mathematical interest. The Hawking radiations from these two kinds of black holes have been studied intensively. Zhang \emph{et al}.~\cite{zhangzhao1,zhangzhao2} and Jiang \emph{et al}.~\cite{jwcprd} studied massless particles tunneling from Kerr black hole and that of charged
particles from the Kerr-Newman black hole in the Parikh-Wilczek tunneling framework. Kerner \emph{et al}.~\cite{kern2}, Li \emph{et al}.~\cite{lrcqg} and Jian \emph{et al}.~\cite{jian} investigate the tunneling of fermions from the Kerr and Kerr-Newman black holes by applying the WKB approximation to the Dirac equation. Umetsu~\cite{Umetsu:2009ra,Umetsu:2010kw} studied the tunneling mechanism in Kerr-Newman black hole by introducing the technique of the dimensional reduction near the horizon. Chen \emph{et al}.~\cite{Chen:2011mg} used the Landauer transport model and one-dimensional quantum channel theory to investigate Hawking radiation of photons and fermions in Kerr and Kerr-Newman black holes. Vieira \emph{et al}.~\cite{Vieira:2014waa} and Konoplya \emph{et al}.~\cite{Konoplya:2014sna} precisely studied the Hawking radiation of charged massive scalar particles from the Kerr-Newman black hole by analytical and numerical methods respectively. All the literature above studied the Hawking radiation from Kerr and Kerr-Newman black holes by investigating the tunneling of photons, scalar or fermion particles.

In this paper, we aim to work out the Hawing radiation of massive vector particles from the Kerr and Kerr-Newman black holes. Vector particles (spin-1 bosons) such as $Z$ and $W^{\pm}$ bosons are well-known and play very important role in Standard Model. When we consider the massive bosons in Kerr space-time or the uncharged massive bosons in Kerr-Newman space-time, the motion of the boson field can be depicted by the Proca equation. In this case, we can study the tunneling mechanism by applying the Hamilton-Jacobi ansatz and WKB approximation to the Proca equation in Kerr or Kerr-Newman space-time, along the same path as in~\cite{Kruglov:2014iya,sik,grch1,grch2,grch3,Sakalli:2015taa}. However, the motion of the charged massive bosons in the Kerr-Newman black hole should be more complicated than the Proca equation because of the nontrivial interaction between the charged massive boson ($W^{\pm}$) fields and the electromagnetic field. Firstly, we derive the field equation of $W$-boson from the Lagrangian given by the Glashow-Weinberg-Salam model. Then we apply the Hamilton-Jacobi ansatz and WKB approximation to the derived equation in Kerr-Newman space-time. By setting the determinant of the derived coefficient matrix to zero, a set of linear equations can be solved for the radial function.  Consequently, we compute the tunneling rate of the charged massive vector particles from the Kerr-Newman black hole and recover the corresponding Hawking temperature.

The remainders of this paper are outlined as follows. In Sec.~\ref{sec:Kerr}, taking
the rotation effect into account, we investigate the Hawking radiation of
vector particles from the Kerr black hole and recover the Hawking temperature.
Extending this work to the charged rotating space-time, both the uncharged and charged bosons tunneling from the Kerr-Newman black hole
are studied in Sec.~\ref{sec:KN}. Sec.~\ref{sec:conclusion} contains some discussion and
conclusion.

\section{Quantum tunneling of massive vector particles from Kerr black hole} \label{sec:Kerr}
In this section, we investigate the Hawking radiation of massive spin-1 particles from rotating (Kerr) black hole. The line element within Kerr space-time is given by
\begin{eqnarray} \label{metric1}
ds^2&=&-(1-\frac{2Mr}{\rho^2})dt^2+\frac{\rho^2}{\Delta}dr^2+\rho^2d\theta^2\notag \\
&&+\left[(r^2+a^2)+\frac{2Mra^2{\rm sin}^2\theta}{\rho^2}\right]{\rm sin}^2\theta d\varphi^2-\frac{4Mra{\rm sin}^2\theta}{\rho^2}dtd\varphi,
\end{eqnarray}
where $\rho^2=r^2+a^2{\rm cos}^2\theta$, $\Delta=r^2-2Mr+a^2$ with $M$ as the black hole mass and $a$ as the angular momentum per unit mass. To make the event horizon
coincide with the infinite red-shift surface, we introduce a new coordinate $\chi=\varphi-\Omega t$ with $\Omega=\frac{2Mra}{(r^2+a^2)^2-\Delta a^2{\rm sin}^2\theta}$, under which the metric \eqref{metric1} becomes
\begin{equation} \label{metric2}
ds^2=-\frac{\Delta \rho^2}{\Sigma(r,\theta)}dt^2+\frac{\rho^2}{\Delta}dr^2+{\rho}^2d\theta^2+\frac{\Sigma(r,\theta)}{\rho^2}{\rm sin}^2\theta d\chi^2,
\end{equation}
where $\Sigma(r,\theta)=(r^2+a^2)^2-\Delta a^2{\rm sin}^2\theta$.

In a curved space-time without electromagnetic field, the motion of massive spin-1 vector field is depicted by the Proca equation
\begin{equation}\label{proca1}
D_\mu\Psi^{\mu\nu}+\frac{m^2}{\hbar^2}\Psi^{\nu}=0,
\end{equation}
in which
\begin{equation}%\label{}
\Psi_{\mu\nu}=D_{\mu}\Psi_{\nu}-D_{\nu}\Psi_{\mu}=\partial_{\mu}\Psi_{\nu}-\partial_{\nu}\Psi_{\mu}.
\end{equation}
Because of the anti-symmetry of $\Psi^{\mu\nu}$ tensor, the Proca equation \label{proca1} is equivalent to
\begin{equation}\label{proca2}
\frac{1}{\sqrt{-g}}\partial_{\mu}(\sqrt{-g}\Psi^{\mu\nu})+\frac{m^2}{\hbar^2}\Psi^{\nu}=0.
\end{equation}

By substituting the metric \eqref{metric2} into \eqref{proca2}, the Proca equation can be reformulated as a tetrad

\begin{eqnarray}
\label{tetrad1}&&\partial_1\left[\frac{\sqrt{-g}\Sigma}{\rho^4}(\partial_0\Psi_1-\partial_1\Psi_0)\right]+\partial_2\left[\frac{\sqrt{-g}\Sigma}{\Delta\rho^4}(\partial_0\Psi_2-\partial_2\Psi_0)\right] \notag\\
&&+\partial_3\left[\frac{\sqrt{-g}}{\Delta{\rm sin}^2\theta}(\partial_0\Psi_3-\partial_3\Psi_0)\right]-\frac{\sqrt{-g}\Sigma m^2}{\Delta \rho^2 \hbar^2}\Psi_0= 0, \\
\label{tetrad2}&&\partial_0\left[\frac{\sqrt{-g}\Sigma}{\rho^4}(\partial_1\Psi_0-\partial_0\Psi_1)\right]-\partial_2\left[\frac{\sqrt{-g}\Delta}{\rho^4}(\partial_1\Psi_2-\partial_2\Psi_1)\right] \notag\\
&&-\partial_3\left[\frac{\sqrt{-g}\Delta}{\Sigma{\rm sin}^2\theta}(\partial_1\Psi_3-\partial_3\Psi_1)\right]+\frac{\sqrt{-g}\Delta m^2}{\rho^2\hbar^2}\Psi_1= 0, \\
\label{tetrad3}&&\partial_0\left[\frac{\sqrt{-g}\Sigma}{\Delta\rho^4}(\partial_2\Psi_0-\partial_0\Psi_2)\right]-\partial_1\left[\frac{\sqrt{-g}\Delta}{\rho^4}(\partial_2\Psi_1-\partial_1\Psi_2)\right] \notag\\
&&-\partial_3\left[\frac{\sqrt{-g}}{\Sigma{\rm sin}^2\theta}(\partial_2\Psi_3-\partial_3\Psi_2)\right]+\frac{\sqrt{-g}m^2}{\rho^2\hbar^2}\Psi_2= 0, \\
\label{tetrad4}&&\partial_0\left[\frac{\sqrt{-g}}{\Delta{\rm sin}^2\theta}(\partial_3\Psi_0-\partial_0\Psi_3)\right]-\partial_1\left[\frac{\sqrt{-g}\Delta}{\Sigma{\rm sin}^2\theta}(\partial_3\Psi_1-\partial_1\Psi_3)\right] \notag\\
&&-\partial_2\left[\frac{\sqrt{-g}}{\Sigma{\rm sin}^2\theta}(\partial_3\Psi_2-\partial_2\Psi_3)\right]+\frac{\sqrt{-g} \rho^2  m^2}{\Sigma{\rm sin}^2\theta\hbar^2}\Psi_3= 0.
\end{eqnarray}

According to the WKB approximation, $\Psi_\mu$ is of the form
\begin{equation} \label{WKB}
\Psi_\mu=C_\mu(t,r,\theta,\chi) {\rm exp}\left[\frac{i}{\hbar}S(t,r,\theta,\chi)\right],
\end{equation}
where $S$ is defined as
\begin{equation} \label{S0123}
    S(t,r,\theta,\chi)=S_0(t,r,\theta,\chi)+\hbar S_1(t,r,\theta,\chi)+\hbar^2 S_2(t,r,\theta,\chi)+\cdots.
\end{equation}
Substituting Eqs. \eqref{WKB}, \eqref{S0123} into the Proca tetrad \eqref{tetrad1}-\eqref{tetrad4} and keeping only the lowest order in $\hbar$, we get equations of the coefficients $C_\alpha$
\begin{eqnarray}
\label{ceq1}&&\frac{\Delta}{\rho^2}\left[C_0(\partial_rS_0)^2-C_1(\partial_rS_0)(\partial_tS_0)\right]+\frac{1}{\rho^2}\left[C_0(\partial_{\theta}S_0)^2-C_2(\partial_{\theta}S_0)(\partial_tS_0)\right] \notag\\
&&+\frac{\rho^2}{\Sigma{\rm sin}^2\theta}\left[C_0(\partial_{\chi}S_0)^2-C_3(\partial_{\chi}S_0)(\partial_tS_0)\right]-C_0m^2= 0, \\
\label{ceq2}&&\frac{-\Sigma}{\Delta\rho^2}\left[C_1(\partial_tS_0)^2-C_0(\partial_tS_0)(\partial_rS_0)\right]+\frac{1}{\rho^2}\left[C_1(\partial_{\theta}S_0)^2-C_2(\partial_{\theta}S_0)(\partial_rS_0)\right] \notag\\
&&+\frac{\rho^2}{\Sigma{\rm sin}^2\theta}\left[C_1(\partial_{\chi}S_0)^2-C_3(\partial_{\chi}S_0)(\partial_rS_0)\right]-C_1m^2= 0, \\
\label{ceq3}&&\frac{-\Sigma}{\Delta\rho^2}\left[C_2(\partial_tS_0)^2-C_0(\partial_tS_0)(\partial_{\theta}S_0)\right]+\frac{\Delta}{\rho^2}\left[C_2(\partial_tS_0)^2-C_1(\partial_rS_0)(\partial_{\theta}S_0)\right] \notag\\
&&+\frac{\rho^2}{\Sigma{\rm sin}^2\theta}\left[C_2(\partial_{\chi}S_0)^2-C_3(\partial_{\chi}S_0)(\partial_{\theta}S_0)\right]-C_2m^2= 0, \\
\label{ceq4}&&\frac{-\Sigma}{\Delta\rho^2}\left[C_3(\partial_tS_0)^2-C_0(\partial_tS_0)(\partial_{\chi}S_0)\right]+\frac{\Delta}{\rho^2}\left[C_1(\partial_tS_0)^2-C_1(\partial_rS_0)(\partial_{\chi}S_0)\right] \notag\\
&&+\frac{1}{\rho^2}\left[C_3(\partial_{\theta}S_0)^2-C_2(\partial_{\theta}S_0)(\partial_{\chi}S_0)\right]-C_3m^2= 0.
\end{eqnarray}

Considering the properties of the Kerr space-time, we carry out separation of variables as
\begin{eqnarray}\label{s0fj}
S_0 &&= -Et+W(r)+j\varphi+\Theta(\theta) \notag\\
&&= -(E-j\Omega)t+W(r)+j\chi+ \Theta(\theta),
\end{eqnarray}
where $E$ and $j$ denote the energy and angular momentum of the emitted particle respectively. Inserting Eq.~\eqref{s0fj} into Eqs.~\eqref{ceq1}-\eqref{ceq4}, one can obtain a matrix equation $K\left(C_{0},C_{1},C_{2},C_{3}\right)^T=0$ and the elements of $K$ are expressed as
\begin{eqnarray}
&&K_{11}= \frac{\Delta W'^2}{\rho^2}+\frac{{J_{\theta}}^2}{\rho^2}+\frac{j^2\rho^2}{\Sigma {\rm sin}^2\theta}-m^2,\ K_{12}=\frac{\Delta W'(E-j\Omega)}{\rho^2},\notag\\
&&K_{13}=\frac{J_{\theta}(E-j\Omega)}{\rho^2},\ K_{14}=\frac{\rho^2j(E-j\Omega)}{\Sigma {\rm sin}^2\theta},\ K_{21}=\frac{-\Sigma W'(E-j\Omega)}{\Delta\rho^2}, \notag \\
&&K_{22}=\frac{\Sigma(E-j\Omega)^2}{\Delta\rho^2}+\frac{{J_{\theta}}^2}{\rho^2}+\frac{j^2\rho^2}{\Sigma {\rm sin}^2\theta}-m^2,\ K_{23}=\frac{-J_{\theta}W'}{\rho^2},\notag\\
&&K_{24}=\frac{-\rho^2jW'}{\Sigma{\rm sin}^2\theta},\ K_{31}=\frac{-\Sigma J_{\theta}(E-j\Omega)}{\Delta \rho^2},\ K_{32}=\frac{-\Delta J_{\theta}W'}{\rho^2}, \\
&&K_{33}=\frac{-\Sigma (E-j\Omega)^2}{\Delta \rho^2}+\frac{\Delta{W'}^2}{\rho^2}+\frac{\rho^2j^2}{\Sigma {\rm sin}^2\theta}-m^2,\ K_{34}=\frac{-\rho^2jJ_{\theta}}{\Sigma {\rm sin}^2\theta}, \notag\\
&&K_{41}=\frac{-\Sigma j(E-j\Omega)}{\Delta \rho^2},\ K_{42}=\frac{-\Delta jW'}{\rho^2},\ K_{43}=\frac{-J_{\theta}j}{\rho^2}, \notag\\
&&K_{44}=\frac{-\Sigma (E-j\Omega)^2}{\Delta \rho^2}+\frac{\Delta{W'}^2}{\rho^2}+\frac{{J_{\theta}}^2}{\rho^2}-m^2, \notag
\end{eqnarray}
where $J_{\theta}$ is identified as $\partial_{\theta}S_{0}$.

The determination of the coefficient matrix should be equal to zero to ensure that Eqs. \eqref{ceq1}-\eqref{ceq4} possess nontrivial solution.  By solving ${\rm det} K=0$, we obtain
\begin{equation}
{W'}^2=\frac{\rho^2{\rm sin}^2\theta\Sigma^2(E-j\Omega)^2+\rho^4{\rm sin}^2\theta m^2\Delta\Sigma-\rho^2{\rm sin}^2\theta{J_{\theta}}^2\Delta\Sigma-\rho^6j^2\Delta}{\Delta^2\rho^2{\rm sin}^2\theta\Sigma},
\end{equation}
which means that there are two values contrary to each other that can be identified as the radial derivatives of the action $S_0$. Correspondingly, we have
\begin{equation}%\label{}
W_{\pm}(r)=\pm\int\sqrt{\frac{\rho^2{\rm sin}^2\theta\Sigma^2(E-j\Omega)^2+\rho^4{\rm sin}^2\theta m^2\Delta\Sigma-\rho^2{\rm sin}^2\theta{J_{\theta}}^2\Delta\Sigma-\rho^6j^2\Delta}{\Delta^2\rho^2{\rm sin}^2\theta\Sigma}}dr.
\end{equation}
We mention that $W_{+}$ denotes the radial function of the outgoing particles and $W_{-}$ of the ingoing particles. Integrating around the pole at the
outer horizon $r_{+}=M+\sqrt{M^2-a^2}$, we obtain
\begin{equation}
W_{\pm}(r)=\pm i\pi\frac{({r_{+}}^2+a^2)(E-j\Omega_{+})}{2(r_{+}-M)},
\end{equation}
where $\Omega_{+}$ is the value of $\Omega$ corresponding to $r=r_{+}$.

So the tunneling probability of the vector particles is
\begin{eqnarray}%\label{}
\Gamma&=&\frac{P_{outgoing}}{P_{ingoing}}=\frac{{\rm exp}\left[-\frac{2}{\hbar}({\rm Im} W_{+}+{\rm Im} \Theta)\right]}{{\rm exp}\left[-\frac{2}{\hbar}({\rm Im} W_{-}+{\rm Im} \Theta)\right]}={\rm exp}\left[-\frac{4}{\hbar}{\rm Im} W_{+}\right] \notag\\
&=&{\rm exp}\left[-\frac{2\pi}{\hbar}\frac{{r_{+}}^2+a^2}{r_{+}-M}(E-j\Omega_{+})\right].
\end{eqnarray}
Setting $\hbar=1$, then Hawking temperature of the Kerr black hole is recovered as
\begin{equation}
T_{H}=\frac{1}{2\pi}\frac{r_{+}-M}{{r_{+}}^2+a^2},
\end{equation}
which is exactly the result obtained by other methods. This implies that when black hole radiates scalar particle, Dirac particles and vector particles, their
tunneling probability and Hawking temperature are the same and are not
related to the kind of particles. The Hawking radiation spectrum of the bosonic particles from Kerr black hole can be deduced following the standard arguments~\cite{tdr,SSGRG}
\begin{equation}\label{hrspectrum}
N(E,j)=\frac{1}{e^{\frac{E-j\Omega_{+}}{K_{B}T_{H}}}-1}.
\end{equation}

\section{Quantum tunneling of massive vector particles from Kerr-Newman black hole} \label{sec:KN}
We can obtain the metric of Kerr-Newman space-time by replacing $2Mr$ in Kerr space-time \eqref{metric1} with $2Mr-Q^2$
\begin{eqnarray} \label{KNmetric1}
ds^2&=&-(1-\frac{2Mr-Q^2}{\rho^2})dt^2+\frac{\rho^2}{\widetilde{\Delta}}dr^2+\rho^2d\theta^2-\frac{2(2Mr-Q^2)a{\rm sin}^2\theta}{\rho^2}dtd\varphi\notag \\
&&+\left[(r^2+a^2)+\frac{(2Mr-Q^2)a^2{\rm sin}^2\theta}{\rho^2}\right]{\rm sin}^2\theta d\varphi^2,
\end{eqnarray}
with the electromagnetic potential
\begin{equation}\label{KNpotential1}
\widehat{A}_{\mu}=\widehat{A}_{t}dt+\widehat{A}_{\varphi}d\varphi=\frac{Qr}{r^2+a^2{\rm cos}^2\theta}dt-\frac{Qra{\rm sin}^2\theta}{r^2+a^2{\rm cos}^2\theta}d\varphi,
\end{equation}
where $Q$ is the charge of the black hole, and $\widetilde{\Delta}=r^2-2Mr+a^2+Q^2$.  Similar to the Kerr black hole case, we introduce coordinate transformation $\chi=\varphi-\widetilde{\Omega}t$, where $\widetilde{\Omega}=\frac{(2Mr-Q^2)a}{(r^2+a^2)^2-\widetilde{\Delta} a^2{\rm sin}^2\theta}$, then the metric \eqref{KNmetric1} becomes
\begin{equation}\label{KNmetric2}
 ds^2=-\frac{\widetilde{\Delta}\rho^2}{\widetilde{\Sigma}(r,\theta)}dt^2+\frac{\rho^2}{\widetilde{\Delta}}dr^2+\rho^2d\theta^2+\frac{\widetilde{\Sigma}(r,\theta)}{\rho^2}{\rm sin}^2\theta d\chi^2,
\end{equation}
with the corresponding potential
\begin{equation}\label{KNpotential2}
A_{\mu}=A_{t}dt+A_{\chi}d\chi=\frac{Qr(r^2+a^2)}{\widetilde{\Sigma}(r,\theta)}dt-\frac{Qra{\rm sin}^2\theta}{r^2+a^2{\rm cos}^2\theta}d\chi,
\end{equation}
where $\widetilde{\Sigma}(r,\theta)=(r^2+a^2)^2-{\widetilde{\Delta}}a{\rm sin}^2\theta$.

The charged bosons ($W^{\pm}$) behave differently from the uncharged bosons ($Z$) in the electromagnetic field. So it's reasonable to investigate their tunneling processes separately.

\subsection{Uncharged massive vector particles} \label{subsec:KNz}
The motion of uncharged massive vector field in the Kerr-Newman background is still depicted by the Proca equation~\eqref{proca2}, except that the metric refers to~\eqref{KNmetric2} now. Because of the similarity of the Kerr-Newman metric~\eqref{KNmetric2} with the Kerr metric~\eqref{metric2}, it should be trivial to extend the study from Kerr case to Kerr-Newman case. One can repeat the process just by replacing $\Delta$, $\Omega$ and $\Sigma$ with $\widetilde{\Delta}$, $\widetilde{\Omega}$ and $\widetilde{\Sigma}$ respectively. As a result, the tunneling probability is
\begin{equation}
\Gamma={\rm exp}\left[-2\pi \frac{{\widetilde{r}_{+}}^2+a^2}{\widetilde{r}_{+}-M}(E-j\widetilde{\Omega}_{+})\right],
\end{equation}
and the Hawking temperature of the Kerr-Newman black hole is recovered as
\begin{equation} \label{KNHT}
T_{H}=\frac{1}{2\pi}\frac{\widetilde{r}_{+}-M}{{\widetilde{r}_{+}}^2+a^2},
\end{equation}
where $\hbar=1$ has been assumed, $\widetilde{r}_{+}=M+\sqrt{M^2-Q^2-a^2}$ is the outer event horizon of Kerr-Newman black hole, and $\widetilde{\Omega}_{+}$ is the value of $\widetilde{\Omega}$ at $r=\widetilde{r}_{+}$.
Again the Hawking temperature we obtained in~\eqref{KNHT} is fully in consistence with that obtained by other methods.
\subsection{Charged massive vector particles} \label{subsec:KNw}
According to the Glashow-Weinberg-Salam model~\cite{GWSmodel}, the Lagrangian of the $W$-bosons in a background electromagnetic field is of the form
\begin{eqnarray}\label{wlag}
\mathcal{L} &=& -\frac{1}{2}\left(D^{+}_{\mu}W^{+}_{\nu}-D^{+}_{\nu}W^{+}_{\mu}\right)\left(D^{-\mu}W^{-\nu}-D^{-\nu}W^{-\mu}\right)+\frac{m_{W}^2}{{\hbar}^2}W^{+}_{\mu}W^{-\mu} \notag \\
 &&-\frac{i}{\hbar}eF^{\mu\nu}W^{+}_{\mu}W^{-}_{\nu},
\end{eqnarray}
where $D^{\pm}_{\mu}=\nabla_{\mu}\pm\frac{i}{\hbar}eA_{\mu}$ with $\nabla_{\mu}$ as the geometrically covariant derivative and $A_{\mu}$ as the electromagnetic potential of the black hole, $e$ denotes the charge of the $W^{+}$ boson, and $F^{\mu\nu}=\nabla^{\mu}A^{\nu}-\nabla^{\nu}A^{\mu}$. The equation of motion of the W-boson field can be derived from Eq.~\eqref{wlag}
\begin{eqnarray}\label{weom}
&&\frac{1}{\sqrt{-g}}\partial_{\mu}\left[\sqrt{-g}\left(D^{\pm\mu}W^{\pm\nu}-D^{\pm\nu}W^{\pm\mu}\right)\right]+\frac{m_{W}^2}{{\hbar}^2}W^{\pm\nu}  \notag\\
&&\pm\frac{i}{\hbar}eA_{\mu}\left(D^{\pm\mu}W^{\pm\nu}-D^{\pm\nu}W^{\pm\mu}\right)\pm\frac{i}{\hbar}eF^{\nu\mu}W^{\pm}_{\mu}=0.
\end{eqnarray}

Since the equation of motion of the $W^{+}$ boson is similar to that of the $W^{-}$ boson, they should share similar tunneling processes too. For simplicity, we will study in detail only the $W^{+}$ boson case, after which the results can be extended to $W^{-}$ boson. Because of the diagonalization of the metric~\eqref{KNmetric2}, the equations of motion of  $W^{+}$ field can be reformulated as
\begin{eqnarray} \label{Wproca}
&&\frac{1}{\sqrt{-g}}\partial_{\mu}\left[\frac{\sqrt{-g}}{g_{\mu\mu}g_{\nu\nu}}\left(\partial_{\mu}W^{+}_{\nu}-\partial_{\nu}W^{+}_{\mu}+\frac{i}{\hbar}e A_{\mu}W^{+}_{\nu}-\frac{i}{\hbar}e A_{\nu}W^{+}_{\mu}\right)\right] \notag\\
&&+\frac{i e A_{\mu}}{\hbar g_{\mu\mu}g_{\nu\nu}}\left(\partial_{\mu}W^{+}_{\nu}-\partial_{\nu}W^{+}_{\mu}+\frac{i}{\hbar}e A_{\mu}W^{+}_{\nu}-\frac{i}{\hbar}e A_{\nu}W^{+}_{\mu}\right) \notag\\
&&+\frac{m_{W}^2}{g_{\nu\nu}\hbar^2}W^{+}_{\nu}+\frac{i}{\hbar}eF^{\nu\mu}W^{+}_{\mu}=0,\ for\ \nu=0,1,2,3,
\end{eqnarray}
where only the summation over $\mu$ is implemented, and the relation $\nabla_{\mu}W^{+}_{\nu}-\nabla_{\nu}W^{+}_{\mu}=\partial_{\mu}W^{+}_{\nu}-\partial_{\nu}W^{+}_{\mu}$ has been used.

According to the WKB approximation, $W^{+}_\mu$ is of the form
\begin{equation} \label{WKBW}
W^{+}_\mu=C_\mu(t,r,\theta,\chi) {\rm exp}\left[\frac{i}{\hbar}S(t,r,\theta,\chi)\right],
\end{equation}
where $S$ is defined as in Eq.~\eqref{S0123}. Substituting Eq.~\eqref{WKBW} and~\eqref{S0123} in Eq.~\eqref{Wproca}, the equations to the leading order in $\hbar$ are
\begin{eqnarray}
\label{Wproca1}&&\frac{\widetilde{\Delta}}{\rho^2}\left[C_0(\partial_rS_0)^2-C_1(\partial_rS_0)(\partial_tS_0+eA_{t})\right]-C_0m_{W}^2 \notag\\
&&+\frac{1}{\rho^2}\left[C_0(\partial_{\theta}S_0)^2-C_2(\partial_{\theta}S_0)(\partial_tS_0+eA_{t})\right] \notag\\
&&+\frac{\rho^2}{\widetilde{\Sigma}{\rm sin}^2\theta}\left[C_0(\partial_{\chi}S_0+eA_{\chi})^2-C_3(\partial_{\chi}S_0+eA_{\chi})(\partial_tS_0+eA_{t})\right]= 0, \\
\label{Wproca2}&&\frac{-\widetilde{\Sigma}}{\widetilde{\Delta}\rho^2}\left[C_1(\partial_tS_0+eA_{t})^2-C_0(\partial_tS_0+eA_{t})(\partial_rS_0)\right]-C_1m_{W}^2 \notag\\
&&+\frac{1}{\rho^2}\left[C_1(\partial_{\theta}S_0)^2-C_2(\partial_{\theta}S_0)(\partial_rS_0)\right] \notag\\
&&+\frac{\rho^2}{\widetilde{\Sigma}{\rm sin}^2\theta}\left[C_1(\partial_{\chi}S_0+eA_{\chi})^2-C_3(\partial_{\chi}S_0+eA_{\chi})(\partial_rS_0)\right]= 0, \\
\label{Wproca3}&&\frac{-\widetilde{\Sigma}}{\widetilde{\Delta}\rho^2}\left[C_2(\partial_tS_0+eA_{t})^2-C_0(\partial_tS_0+eA_{t})(\partial_{\theta}S_0)\right]-C_2m_{W}^2 \notag\\
&&+\frac{\widetilde{\Delta}}{\rho^2}\left[C_2(\partial_tS_0+eA_{t})^2-C_1(\partial_rS_0)(\partial_{\theta}S_0)\right] \notag\\
&&+\frac{\rho^2}{\widetilde{\Sigma}{\rm sin}^2\theta}\left[C_2(\partial_{\chi}S_0+eA_{\chi})^2-C_3(\partial_{\chi}S_0+eA_{\chi})(\partial_{\theta}S_0)\right]= 0, \\
\label{Wproca4}&&\frac{-\widetilde{\widetilde{\Sigma}}}{\widetilde{\Delta}\rho^2}\left[C_3(\partial_tS_0+eA_{t})^2-C_0(\partial_tS_0+eA_{t})(\partial_{\chi}S_0+eA_{\chi})\right]-C_3m_{W}^2\notag\\
&&+\frac{\widetilde{\Delta}}{\rho^2}\left[C_1(\partial_tS_0+eA_{t})^2-C_1(\partial_rS_0)(\partial_{\chi}S_0+eA_{\chi})\right] \notag\\
&&+\frac{1}{\rho^2}\left[C_3(\partial_{\theta}S_0)^2-C_2(\partial_{\theta}S_0)(\partial_{\chi}S_0+eA_{\chi})\right]= 0.
\end{eqnarray}
Note that the nontrivial interaction term $\pm\frac{i}{\hbar}eF^{\nu\mu}W^{\pm}_{\mu}$ in Eq.~\eqref{weom} doesn't contribute to the motion of the $W$-boson field at the leading order in $\hbar$.

Considering the properties of the Kerr-Newman space-time, $S_0$ exists with the same form as in the Kerr space-time
\begin{eqnarray}\label{KNs0fj}
S_0 &&= -Et+W(r)+j\varphi+\Theta(\theta) \notag\\
&&= -(E-j\widetilde{\Omega})t+W(r)+j\chi+ \Theta(\theta).
\end{eqnarray}
Inserting Eq.~\eqref{KNs0fj} into Eqs.~\eqref{Wproca1}-\eqref{Wproca4}, one can obtain a matrix equation
\begin{equation}\label{matrixeq}
K(C_{0}, C_{1}, C_{2}, C_{3})^{T}=0,
\end{equation}
where $K$ is a 4$\times$4 matrix whose elements are
\begin{eqnarray}
&&K_{11}= \frac{\widetilde{\Delta} W'^2}{\rho^2}+\frac{{J_{\theta}}^2}{\rho^2}+\frac{(j+eA_{\chi})^2\rho^2}{\widetilde{\Sigma} {\rm sin}^2\theta}-m_{W}^2, \notag\\
&&K_{22}=\frac{\widetilde{\Sigma}(E-eA_{t}-j\widetilde{\Omega})^2}{\widetilde{\Delta}\rho^2}+\frac{{J_{\theta}}^2}{\rho^2}+\frac{(j+eA_{\chi})^2\rho^2}{\widetilde{\Sigma} {\rm sin}^2\theta}-m_{W}^2, \notag \\
&&K_{33}=\frac{-\widetilde{\Sigma} (E-eA_{t}-j\widetilde{\Omega})^2}{\widetilde{\Delta} \rho^2}+\frac{\widetilde{\Delta}{W'}^2}{\rho^2}+\frac{\rho^2(j+eA_{\chi})^2}{\widetilde{\Sigma} {\rm sin}^2\theta}-m_{W}^2, \notag \\
&&K_{44}=\frac{-\widetilde{\Sigma} (E-eA_{t}-j\widetilde{\Omega})^2}{\widetilde{\Delta} \rho^2}+\frac{\widetilde{\Delta}{W'}^2}{\rho^2}+\frac{{J_{\theta}}^2}{\rho^2}-m_{W}^2, \notag \\
&&K_{12}=\frac{\widetilde{\Delta} W'(E-eA_{t}-j\widetilde{\Omega})}{\rho^2},\ K_{13}=\frac{J_{\theta}(E-eA_{t}-j\widetilde{\Omega})}{\rho^2},  \\
&&K_{14}=\frac{\rho^2(j+eA_{\chi})(E-eA_{t}-j\widetilde{\Omega})}{\widetilde{\Sigma} {\rm sin}^2\theta},\ K_{21}=\frac{-\widetilde{\Sigma} W'(E-eA_{t}-j\widetilde{\Omega})}{\widetilde{\Delta}\rho^2}, \notag \\ &&K_{23}=\frac{-J_{\theta}W'}{\rho^2},\ K_{24}=\frac{-\rho^2(j+eA_{\chi})W'}{\widetilde{\Sigma}{\rm sin}^2\theta},\notag\\
&&K_{31}=\frac{-\widetilde{\Sigma} J_{\theta}(E-eA_{t}-j\widetilde{\Omega})}{\widetilde{\Delta} \rho^2},\ K_{32}=\frac{-\widetilde{\Delta} J_{\theta}W'}{\rho^2}, \notag \\
&&K_{34}=\frac{-\rho^2(j+eA_{\chi})J_{\theta}}{\widetilde{\Sigma} {\rm sin}^2\theta},\ K_{41}=\frac{-\widetilde{\Sigma} (j+eA_{\chi})(E-eA_{t}-j\widetilde{\Omega})}{\widetilde{\Delta} \rho^2}, \notag\\
&&K_{42}=\frac{-\widetilde{\Delta} (j+eA_{\chi})W'}{\rho^2},\ K_{43}=\frac{-J_{\theta}(j+eA_{\chi})}{\rho^2}. \notag
\end{eqnarray}

Eq.~\eqref{matrixeq} possesses nontrivial solution if the determinant of the matrix $K$ equals zero. By solving ${\rm det} K=0$, we obtain
\begin{equation}%\label{}
{W'}^2=\frac{\rho^2{\rm sin}^2\theta\widetilde{\Sigma}^2(E-eA_{t}-j\widetilde{\Omega})^2+\rho^4{\rm sin}^2\theta m_{W}^2\widetilde{\Delta}\widetilde{\Sigma}-\rho^2{\rm sin}^2\theta{J_{\theta}}^2\widetilde{\Delta}\widetilde{\Sigma}-\rho^6(j+eA_{\chi})^2\widetilde{\Delta}}{{\widetilde{\Delta}}^2\rho^2{\rm sin}^2\theta\widetilde{\Sigma}},
\end{equation}
Integrating $\pm W'$ around the pole at the outer horizon $\widetilde{r}_{+}=M+\sqrt{M^2-Q^2-a^2}$, we obtain
\begin{equation}
W_{\pm}(r)=\pm i\pi\frac{({\widetilde{r}_{+}}^2+a^2)(E-eA_{t+}-j\widetilde{\Omega}_{+})}{2(\widetilde{r}_{+}-M)},
\end{equation}
where $\widetilde{\Omega}_{+}$ and $A_{t+}$ are the values of $\widetilde{\Omega}$ and $A_{t}$ corresponding to $r=\widetilde{r}_{+}$ respectively, $W_{+}$ denotes the radial function of the outgoing particles and $W_{-}$ of the ingoing particles. Then we can obtain the tunneling probability of the $W^{+}$ bosons
\begin{eqnarray}%\label{}
\Gamma(W^{+})&=&{\rm exp}\left[-\frac{4}{\hbar}{\rm Im} W_{+}\right] \notag\\
&=&{\rm exp}\left[-\frac{2\pi}{\hbar}\frac{{\widetilde{r}_{+}}^2+a^2}{\widetilde{r}_{+}-M}(E-eA_{t+}-j\widetilde{\Omega}_{+})\right],
\end{eqnarray}
and Hawking temperature at the outer event horizon of Kerr-Newman black hole is recovered as
\begin{equation}\label{htwb}
T_{H}=\frac{1}{2\pi}\frac{\widetilde{r}_{+}-M}{{\widetilde{r}_{+}}^2+a^2},
\end{equation}
which is fully in consistence with Eq.~\eqref{KNHT}.

The $W^{-}$ boson case proceeds in a manner fully analogous to the $W^{+}$ boson case discussed above. Similarly, the tunneling probability of $W^{-}$ bosons can be obtained as
\begin{equation}%\label{}
\Gamma(W^{-})={\rm exp}\left[-\frac{2\pi}{\hbar}\frac{{\widetilde{r}_{+}}^2+a^2}{\widetilde{r}_{+}-M}(E+eA_{t+}-j\widetilde{\Omega}_{+})\right],
\end{equation}
and the Hawking temperature in Eq.~\eqref{htwb} is recovered once again.

\section{Conclusions and discussions} \label{sec:conclusion}
In this paper, by applying the WKB approximation and the Hamilton-Jacobi ansatz to the massive vector field equations, we have investigated the tunneling mechanism of massive bosons from the Kerr black hole and Kerr-Newman black hole. The expected Hawking temperatures have been recovered and proven to be consistent with black hole universality. We should mention that in our investigation, the back reaction of the emitted particle on the black hole geometry and self-gravitational interaction are reasonably neglected, and the derived Hawking temperatures are only the leading terms. One can compute corrections to the massive bosons tunneling probability and the emission temperature by fully taking into account conservation of energy and charge.

\textbf{Acknowledgments}

X.-Q. Li is supported by the National Nature Science Foundation of China (NSFC) under Grant No. 11475237, No. 11121064 and No. 10821504.

%%%%%%%%%%%%%%%%%%%%%%%%%%%%%%%%%%%%%%%%%%%%

\end{document}